\documentclass[a4paper,11pt]{article}
\pdfoutput=1 % if your are submitting a pdflatex (i.e. if you have
\usepackage{jcappub} % for details on the use of the package, please
\usepackage{epsfig}
\usepackage{epstopdf}
\input{epsf.sty}
\usepackage{epsf}
\usepackage{amssymb}
\usepackage{caption}
\usepackage{subcaption}
\usepackage{tensor}
\usepackage{yhmath}

\def\be{\begin{equation}}
\def\ee{\end{equation}}
\def\ba{\begin{eqnarray}}
\def\ea{\end{eqnarray}}
\include{mydefs}

\newcommand{\dd}{{\rm d}}
\newcommand{\pd}[3][]{\frac{\partial^{#1} #2}{\partial {#3}^{#1}}}

\newcommand{\eps}{\varepsilon}
\newcommand{\define}{\equiv}
\newcommand{\vect}[1]{\boldsymbol{#1}}
\newcommand{\grad}{\vect{\nabla}}
\newcommand{\elec}{\vect{E}}
\newcommand{\magn}{\vect{B}}
\newcommand{\K}{K}
\newcommand{\X}{X}
\newcommand{\Y}{Y}
\newcommand{\Z}{Z}
\newcommand{\Lagr}{\mathcal{L}}
\newcommand{\pFtilde}{\mathcal{B}}

\newcommand{\pa}[1]{\left( #1 \right)}
\newcommand{\pac}[1]{\left[ #1 \right]}
\newcommand{\paac}[1]{\left\{ #1 \right\}}
\newcommand{\abs}[1]{\left| #1 \right|}

\definecolor{blue4}{RGB}{0,0,143}

\title{ 
On the stability and causality of scalar-vector theories
}

\author[a,b]{Pierre Fleury,}
\author[c]{Juan P. Beltr\'an Almeida,}
\author[a,b]{Cyril Pitrou,}
\author[a,b]{Jean-Philippe Uzan.}
\affiliation[a]{Institut d'Astrophysique de Paris, CNRS UMR 7095, Universit\'e Pierre \& Marie Curie - Paris VI, 98 bis Bd Arago, 75014 Paris, France.}
\affiliation[b]{Sorbonne Universit\'es, Institut Lagrange de Paris, 98 bis, Bd Arago, 75014 Paris, France.}
\affiliation[c]{Departamento de F\'isica, Universidad Antonio Nari\~no,  Cra 3 Este \# 47A-15, Bogot\'a DC, Colombia.}

\emailAdd{fleury@iap.fr}
\emailAdd{juanpbeltran@uan.edu.co}
\emailAdd{pitrou@iap.fr}
\emailAdd{uzan@iap.fr}

\abstract{Various extensions of standard inflationary models have been
  proposed recently by adding vector fields. Because they are
  generally motivated by large-scale anomalies, and the possibility of
  statistical anisotropy of primordial fluctuations, such models
  require to introduce non-standard couplings between vector fields on the one hand, and either gravity or scalar fields on the other hand. In this article, we study models involving a vector field coupled to a scalar field. We derive restrictive necessary conditions for these models to be both stable (Hamiltonian bounded by below) and causal (hyperbolic equations of motion).}

 
\begin{document}

\maketitle

%%%%%%%%%%%%%%%%%%%%%%%%%%%%%
\section{Introduction}
%%%%%%%%%%%%%%%%%%%%%%%%%%%%%

Inflationary models including a vector-field sector have been studied following diverse approaches over the
past few years. Among the most recent models, a sizable fraction was motivated
mainly by the appearance of certain ``anomalies" pointed out by
observations~\cite{Groeneboom0, Groeneboom09, Hanson09, Hanson10,
  Bennett12, Hinshaw12}, later confirmed by \textit{Planck}'s results
\cite{PlanckSI13}, which suggested the presence of statistical
anisotropies and maybe signals of parity violation in the cosmic microwave background (CMB). Although the
statistical significance of these anomalies is still a matter of debate~\cite{Kim:2013gka,Ramazanov:2013wea} (possible systematic errors, contamination by foregrounds, asymmetric beams, etc.), it is not excluded that they are
actually seeded by a source, different than an inflaton scalar field,
during the early stages of the Universe. In this context, vector fields
arise as suitable, simple and natural candidates to explain the origin
of such anomalies as they possess intrinsically a preferred direction. With these motivations and the requirement to
generate both the inflationary dynamics and the presence of a detectable level of statistical anisotropy in the CMB
within an unified framework, several models involving vector fields
have recently been proposed. Their classical dynamics and
statistical properties have thus been explored in great details
\cite{Dimopoulos06, Ackerman07, Golovnev08, Dimopoulos08, Yokoyama08,
  Himmetoglu08, Himmetoglu:2008hx, Himmetoglu:2009qi, Karciauskas08,
  Dimopoulos09a, Dimopoulos09vu, EspositoFarese09aj, Watanabe:2009ct,
  Bartolo09a, Bartolo09b, Valenzuela09a, Valenzuela09b, Dulaney10, Gumrukcuoglu10, Watanabe10,
  Maleknejad11, Maleknejad:2011jr, Maleknejad:2011jw, Dimopoulos11,
  Valenzuela11, Bartolo12, Rodriguez13, Abolhasani13, Lyth13,
  Shiraishi13, Namba:2013kia, Adshead:2012kp, Adshead:2013nka,
  Ohashi:2013qba, Namba12gg, Fujita:2013qxa, Maeda:2013daa} (for
reviews see Refs. \cite{Dimastrogiovanni10, Soda12,
  Maleknejad12}). The determination of cosmological
parameters related to the presence of a statistical anisotropy in the CMB can
provide valuable information about the mechanisms governing the dynamics of the inflationary Universe, and their possible deviations from the reference single-field model.

%Statistical parameters such as non-Gaussianity parameters, statistical anisotropies, polarization spectra, birefringence, among others, have been calculated and analyzed in great detail in last years...
%Aside of the interest  for the statistical features in vector field models mentioned before and their potential as measurable quantities, inflationary models with vector fields are interesting on their own because they have a structure and a phenomenology much richer  than those models based only on scalar fields. With these models one also have a rich amount of predictions and one can define new meaningful statistical parameters relating intrinsic parameters of the models with potentially observable statistical parameters  which could eventually be tested with the required amount of precision in the next future...  \\

In the recent literature, popular models propose to couple scalar and vector fields by modifying the standard kinetic term of the vector as $f(\phi)F^{\mu\nu}F_{\mu\nu}$~\cite{Yokoyama08, Watanabe:2009ct}; or adding a term of the form $\phi F_{\mu \nu} \tilde{F}^{\mu \nu}$ ($\tilde{F}$ being the Hodge dual of $F$) which couples vectors and ``pseudo scalars" or axions~\cite{Anber:2006xt, Sorbo:2011rz, Urban:2013spa, Dimopoulos:2012av, Shiraishi:2013kxa} (see the review \cite{Pajer:2013fsa} for further references); or variants and generalizations of the above ideas including non-Abelian gauge fields~\cite{Bartolo09a, Bartolo09b, Dimastrogiovanni10, Adshead:2012kp, Maeda:2013daa}. These models have been proved to be free from instabilities, in particular they do not possess any longitudinal propagating mode; they also have the virtue of generating a non-diluting amount of statistical anisotropy \cite{Watanabe:2009ct, Bartolo12} which could leave measurable imprints in the CMB. Note also that such scalar-vector models have been proposed recently~\cite{Caprini:2014mja} to give an inflationary origin to extragalactic magnetic fields~\cite{2010Sci...328...73N,2011A&A...529A.144T,2012ApJ...747L..14V}.

Of course, scalar-vector theories offer a very broad set of possibilities, among which the examples mentioned above are somehow the simplest representatives. Apart from Occam's razor, there is a priori no reason to focus on these models specifically, hence one could wonder which subset of all possibilities are worth investigating. This motivates the present article, whose purpose is to identify a class of fundamentally healthy scalar-vector theories, which could then be safely considered candidates for inflationary or dark energy models. More precisely, we focus on theories involving one scalar field and one (gauge-invariant) vector field, both minimally coupled to spacetime geometry, and we study the necessary conditions for such theories to be stable---their Hamiltonian must be bounded by below---and causal---their dynamics must be governed by hyperbolic equations of motion. This method has the advantage of being nonperturbative, and thus more general than only studying the behavior of small perturbations about a given background. The application of the healthy models to cosmology, together with the tests of their compatibility with current observations, are beyond the scope of our analysis and left as future projects.

The article is organized as follows. In Sec.~\ref{sec:action}, we derive the most general form of a theory involving a scalar field and a gauge-invariant vector field, both minimally coupled to gravity, and propose a reasonable restriction motivated by previous works. The Hamiltonian stability of this theory is analyzed in Sec.~\ref{stability}, and then its causality in Sec.~\ref{sec:causality}. Finally, Sec.~\ref{rd} is dedicated to a summary of the results, followed by a discussion about possible extensions.

%%%%%%%%%%%%%%%%%%%%%%%%%%%%%%%%%%%%%%%%%%%%%%%%%%%%%%
%%%%%%%%%%%%%%%%%%%%%%%%%%%%%%%%%%%%%%%%%%%%%%%%%%%%%%
\section{Building general scalar-vector models}
\label{sec:action}
%%%%%%%%%%%%%%%%%%%%%%%%%%%%%%%%%%%%%%%%%%%%%%%%%%%%%%
%%%%%%%%%%%%%%%%%%%%%%%%%%%%%%%%%%%%%%%%%%%%%%%%%%%%%%

%%%%%%%%%%%%%%%%%%%%%%%%%%%%%%%%%%%%%%%%%%%%%%%%%%%%%%
\subsection{General assumptions}
%%%%%%%%%%%%%%%%%%%%%%%%%%%%%%%%%%%%%%%%%%%%%%%%%%%%%%

Consider, as a starting point, the most conservative theory in which matter is described by the standard model of particle physics, while being minimally coupled to spacetime geometry governed by general relativity. To this theory, we add two new fields, namely a scalar~$\phi$ and a vector $A^\mu$, which for convenience will be referred to as ``dark sector'', though it can stand for inflationary models as well as for dark energy models (see, e.g., Ref.~\cite{Uzan:2006mf} for a discussion of the different classes of universality of extensions).

The presence of such new fields potentially offers a huge amount of possibilities, depending on how they couple to standard matter, to spacetime geometry, or simply to each other. Among them, many shall lead to unhealthy theories, typically due to instabilities (e.g., ghosts), or violations of causality. Since obviously we cannot analyse all possible theories, we choose to focus on those satisfying the following three conditions.

%In view of constructing, e.g., inflationary models where the vector field sector has a sizable signature, many extensions of the action~\eqref{eq:simplest_action} are possible: couplings between $\phi$ and $A^\mu$, non-minimal coupling of the fields to spacetime geometry, dropping gauge invariance (by introducing, e.g., a mass for the vector field, such as in Proca theory), higher-order derivatives of the fields, etc. Among all the possible models, many are expected to be unhealthy, typically because of the presence of instabilities (e.g., ghosts) or violations of causality. Since it would clearly be impossible to analyse all the possibilities, we choose to focus on the class of models satisfying the following three conditions:

%
\begin{enumerate}
\item \emph{The fields~$\phi$, $A^\mu$ are uncoupled to standard matter, and minimally coupled to gravity}. In other words, the action of the theory takes the form
\begin{equation}
S = S_{\rm EH}[g_{\mu\nu}] + S_{\rm SM}[\psi_{\rm m};g_{\mu\nu}] + S_{\rm DS}[\phi,A^\mu;g_{\mu\nu}],
\end{equation}
where $S_{\rm EH}$ is the Einstein-Hilbert action, $S_{\rm SM}$ the action of the standard model of particle physics, and $S_{\rm DS}$ the action of the dark sector. This assumption ensures (a)~the non-violation of the equivalence principle,
%\footnote{By ``equivalence principle'', we mean that the effects of gravity on $\phi$ and $A^\mu$---if considered as matter fields---can be locally suppressed by working in a free-falling frame. However, an alternative point of view could consist in viewing $\phi$ as making part of gravity (such as in scalar-tensor theories); in the latter case, the equivalence principle is generally violated, because two $A^\mu$-bosons with different energies can fall differently.}
and (b) the constancy of fundamental constants~\cite{Uzan:2002vq,Uzan:2010pm}. Note that for a scalar field alone, non-minimal couplings to spacetime geometry have been actively studied, e.g. in the context of scalar-tensor theories, and now well understood~\cite{Damour:1992we,Bruneton:2007si}. For a vector field alone, it has been proved that non-minimal coupling generically leads to instabilities~\cite{Himmetoglu:2009qi,EspositoFarese09aj}. See also Refs.~\cite{Barrow:2012ay,Jimenez:2013qsa} for stability analyses of Horndeski's vector-tensor theory~\cite{Horndeski:1976gi} in a cosmological context.
\label{hyp:minimal_coupling}
\item \emph{The action only contains at most order-one derivatives of~$\phi$, $A^\mu$.} This is a sufficient condition to have second-order equations of motion, though not necessary---see for instance Horndeski and Galileon models~\cite{Horndeski:1974wa, Nicolis:2008in, Deffayet:2009mn, Deffayet:2013lga,Khoury:2013tda,Heisenberg:2014rta}.
\label{hyp:first-order}
\item \emph{The action is gauge invariant\footnote{In the sense of gauge transformations of the vector field, i.e. $A_\mu\rightarrow A_\mu + \partial_\mu\Lambda$, where $\Lambda$ is an arbitrary scalar function.}}. This restriction is essentially chosen for simplicity. The variety of models breaking gauge invariance is indeed extremely broad, even in the absence of scalar fields (see, e.g., Refs.~\cite{Heisenberg:2014rta, Tasinato:2014eka}), which would make the analysis performed in the present article much more involved.
\label{hyp:gauge_invariance}
\end{enumerate}

The last two asumptions imply that the action of the dark sector reads
\begin{equation}
S_{\rm DS}
=
\int \dd^4 x \sqrt{-g} \; \Lagr_{\rm DS} (\phi,\partial_{\mu}\phi,F_{\mu\nu};g_{\mu\nu}),
\end{equation}
where $g$ is the determinant of spacetime's metric, and $F\define\dd A=(F_{\mu\nu}/2)\,\dd x^\mu\wedge\dd x^\nu$, with $F_{\mu\nu}=\partial_\mu A_\nu-\partial_\nu A_\mu$, is the field-strength two-form associated with the vector field. The latter can only appear through $F$ in $\Lagr_{\rm DS}$, since any other type of term (e.g., $A^\mu A_\mu$ or $\partial_\mu A^\mu$) would be gauge dependent, which is excluded by assumption~\ref{hyp:gauge_invariance}.

%%%%%%%%%%%%%%%%%%%%%%%%%%%%%%%%%%%%%%%%%%%%%%%%%%%%%%
\subsection{A general class of Lagrangians}
\label{subsec:general_class_Lagrangian}
%%%%%%%%%%%%%%%%%%%%%%%%%%%%%%%%%%%%%%%%%%%%%%%%%%%%%%

Let us construct the most general Lagrangian density for the dark sector, under the assumptions formulated above. As a scalar, $\mathcal{L}_{\rm DS}$ can only depend on the scalars that can be constructed from $\phi$, $\partial_\mu\phi$, $F_{\mu\nu}$; in principle, their free indices could be contracted with arbitary tensors---standing for parameters of the theory---and lead to terms of the form
\begin{equation}
T^{\alpha_1\ldots\alpha_n \mu_1\ldots\mu_p\nu_1\ldots\nu_p}(\phi)\,
\partial_{\alpha_1}\phi \ldots \partial_{\alpha_n}\phi
F_{\mu_1\nu_1} \ldots F_{\mu_p\nu_p}.
\end{equation}
However, from a tensorial parameter there generally emerges fundamentally preferred directions in spacetime\footnote{An example can be found in Refs.~\cite{Mukohyama:2013ew,Kehayias:2014uta}, where the arrow of time emerges from the gradient of a nondynamical scalar within a Riemannian manifold.}, that we do not wish in the theories considered here. The only nondynamical tensor escaping from this rule is the Levi-Civita tensor~$\eps_{\mu\nu\rho\sigma}\define-\sqrt{-g}[\mu\nu\rho\sigma]$, where $[\mu\nu\rho\sigma]$ stands for the permutation symbol, with the convention $[0123]=1$. It turns out that any scalar constructed from $\phi$, $\partial_\mu\phi$, $F_{\mu\nu}$, $g_{\mu\nu}$, and $\eps_{\mu\nu\rho\sigma}$ can be written as a function of $\phi$,
\begin{align}
\K &\define \partial_\mu\phi\partial^\mu\phi, \\
\X &\define F^{\mu\nu}F_{\mu\nu}, \\
\Y &\define F^{\mu\nu}\tilde{F}_{\mu\nu}, \\
\Z &\define (\partial_\mu\phi \tilde{F}\indices{^\mu^\alpha}) (\partial_\nu\phi  \tilde{F}\indices{^\nu_\alpha}),
\end{align}
where $\tilde{F}_{\mu\nu}\define\eps_{\mu\nu\rho\sigma} F^{\rho\sigma}/2$ are the components of the Hodge dual~$\star F$ of $F$; so that
\begin{equation}
\mathcal{L}_{\rm DS}(\phi,\K,\X,\Y,\Z).
\label{eq:general_Lagrangian_DS}
\end{equation}

Let us prove this assertion.
%We first note that a well behaved $\mathcal{L}_{\rm matter}$ can always be decomposed as a sum of products  of scalars, possibly an infinite sum for a general analytic function, with each scalar being built from a product of the $\phi$, $\partial_\mu\phi$, $F_{\mu\nu}$, $g_{\mu\nu}$ and $\eps_{\mu\nu\rho\sigma}$ with no free index. Then, for each of these products,
First, it is clear that the Levi-Civita tensor cannot be involved without being contracted with $F_{\mu\nu}$; if both indices of the latter are contracted with the former, then it leads to $\tilde{F}_{\mu\nu}$; if only one index is contracted, then we get
\begin{align}
\eps_{\mu\nu\rho\sigma} F^{\lambda\sigma} 
&= -\frac12 \eps_{\mu\nu\rho\sigma} \eps^{\alpha\beta\lambda\sigma} \tilde{F}_{\alpha\beta} \\
&= \frac12 \delta^{[\alpha}_\mu \delta^\beta_\nu \delta^{\lambda]}_\rho \tilde{F}_{\alpha\beta} \\
&= \tilde{F}_{\mu\nu} \delta^\lambda_\rho + \tilde{F}_{\rho\mu} \delta^\lambda_\nu + \tilde{F}_{\nu\rho} \delta^\lambda_\mu.
\label{eq:epsilon_and_F}
\end{align}
So when the Levi-Civita tensor appears, the associated expression can be rewritten in terms of $\star F$, whence $\mathcal{L}_{\rm DS}(\phi,\partial_\mu\phi,F_{\mu\nu},\tilde{F}_{\mu\nu})$. There are two elementary classes of scalars that can be constructed from contractions of $\partial_\mu\phi,F_{\mu\nu},\tilde{F}_{\mu\nu}$, namely
\begin{equation}
\wideparen{F}\indices{^\mu_{\alpha_1}} 
\wideparen{F}\indices{^{\alpha_1}_{\alpha_2}}
\ldots
\wideparen{F}\indices{^{\alpha_n}_\mu},
\qquad \text{or} \qquad
\partial_\mu \phi 
\pa{ \wideparen{F}\indices{^\mu_{\alpha_1}} 
		\wideparen{F}\indices{^{\alpha_1}_{\alpha_2}}
		\ldots
		 \wideparen{F}\indices{^{\alpha_n}_\nu}
		}
\partial^\nu \phi,
\label{eq:chains}
\end{equation}
where $\wideparen{F}$ stands either for $F$ or for $\tilde F$. Indeed, since $\partial_\mu\phi$ has only one index, it always ends a contraction branch, hence if more than two $\partial_\mu\phi$ are involved in a scalar term, then it can be factorized into chains of the form \eqref{eq:chains}. Finally, such $\wideparen{F}$-chains can in general be reduced thanks to the identities\footnote{These identities can be considered a special case of the following \emph{lemma}: in a four-dimensional manifold, for any two 2-forms $A=(A_{\mu\nu}/2)\dd x^\mu\wedge\dd x^\nu$ and $B=(B_{\mu\nu}/2)\dd x^\mu\wedge\dd x^\nu$,
\begin{equation}
A^{\mu\alpha} \tilde{B}_{\nu\alpha}
+ B^{\mu\alpha} \tilde{A}_{\nu\alpha}
=
\frac{1}{2} \pa{ B^{\alpha\beta} \tilde{A}_{\alpha\beta} }
\delta^\mu_\nu .
\end{equation}
This can be easily derived by using the contraction of two Levi-Civita tensors $\eps_{\mu\nu\rho\sigma} \eps^{\alpha\beta\lambda\sigma}=-\delta^{[\alpha}_\mu \delta^\beta_\nu \delta^{\lambda]}_\rho$.}
% End footnote.
%
\begin{align}
F^{\mu\alpha} F_{\nu\alpha} - \tilde{F}^{\mu\alpha} \tilde{F}_{\nu\alpha}
&
%\frac{1}{2} \pa{ F^{\alpha\beta} F_{\alpha\beta} } \delta^\mu_\nu
= \frac12 \X \delta^\mu_\nu,
\label{eq:identity_FF}
\\
F^{\mu\alpha} \tilde{F}_{\nu\alpha}
&
%\frac{1}{4} \pa{ F^{\alpha\beta} \tilde{F}_{\alpha\beta} } \delta^\mu_\nu 
= \frac14 \Y \delta^\mu_\nu.
\label{eq:identity_FFtilde}
\end{align}
Indeed, if in a $\wideparen{F}$-chain, an $F$ and an $\tilde{F}$ are contiguous, then we can use Eq.~\eqref{eq:identity_FFtilde} to factorize the couple. If there are only $F$s (or only $\tilde{F}$s) in a chain with strictly more than two $\wideparen{F}$s, then we use Eq.~\eqref{eq:identity_FF} to create $F\tilde{F}$ pairs, and so on. The only irreducible chain\footnote{The situation changes, however, in the case of non-Abelian gauge fields, since there appear non-zero terms of the form $FFF$, $FF\tilde{F}$, etc. This will be briefly discussed in Sec.~\ref{rd}.} is therefore $F^{\mu\alpha} F_{\nu\alpha}$ (or alternatively $\tilde{F}^{\mu\alpha} \tilde{F}_{\nu\alpha}$), that is, if contracted with the gradient of the scalar field, $\partial_\mu\phi F^{\mu\alpha} F_{\nu\alpha}\partial^\nu\phi$ (or alternatively $\partial_\mu\phi \tilde{F}^{\mu\alpha} \tilde{F}_{\nu\alpha}\partial^\nu\phi=-\Z$). In this article, we consider $\Z$ instead of the untilded contraction, because it will turn out to be more convenient for presenting the results of Sec.~\ref{stability}.

%%%%%%%%%%%%%%%%%%%%%%%%%%%%%%%%%%%%%%%%%%%%%%%%%%%%%%
\subsection{A reasonable restriction}
%%%%%%%%%%%%%%%%%%%%%%%%%%%%%%%%%%%%%%%%%%%%%%%%%%%%%%

In Ref.~\cite{EspositoFarese09aj}, the authors have analyzed the stability and causality conditions for vector theories whose Lagrangian density is an arbitrary function of $F^2$ and $F\tilde{F}$, i.e. $\mathcal{L}_{\rm vec}(X,Y)$. Although general conclusions could not be drawn, it appeared that nonlinear functions of only $\X$, or only $\Y$, are excluded. This motivates our fourth restrictive assumption: \emph{we only consider models which are at most linear in $\X$, $\Y$, and $\Z$, i.e., at most quadratic in the vector field}. Thus, in the remainder of this article, we consider a dark-sector Lagrangian density of the form
\begin{equation}\label{eq:action}
\Lagr_{\rm DS}
=
-\frac12 f_0(\phi,\K)
- \frac14 f_1(\phi,\K) \X
- \frac14 f_2(\phi,\K) \Y
+ \frac12 f_3(\phi,\K) \Z,
\end{equation}
and investigate under which conditions on the four functions $f_{0,1,2,3}$ a model is both stable---Hamiltonian bounded by below---and causal---hyperbolic equations of motion.

Our analysis can also be considered a starting point for more ambitious ones, where some of our four restrictive assumptions would be dropped (see, e.g., appendix~\ref{app:beyond_linearity_X_Y_Z} for elements about Lagrangian densities $\mathcal{L}_{\rm DS}$ which are nonlinear in $\X$, $\Y$, $\Z$).

%%%%%%%%%%%%%%%%%%%%%%%%%%%%%%%%%
%%%%%%%%%%%%%%%%%%%%%%%%%%%%%%%%%
\section{Stability of the models}
\label{stability}
%%%%%%%%%%%%%%%%%%%%%%%%%%%%%%%%%
%%%%%%%%%%%%%%%%%%%%%%%%%%%%%%%%%

In this section, we turn to the study of the stability of a dark sector defined by the Lagrangian density~\eqref{eq:action}. 
After having computed the associated Hamiltonian density (Subsec.~\ref{subsec:hamiltonian_formulation}), we investigate in details the conditions under which it is bounded by below (Subsec.~\ref{subsec:hamiltonian_stability}), that is necessary for the stability of the quantum theory, and we summarize the results in Subsec.~\ref{subsec:summary_stability}. In this last subsection, we also discuss why all the results, though derived in Minkowski spacetime, are also completely valid in the presence of gravity.

%%%%%%%%%%%%%%%%%%%%%%%%%%%%%%%%%%%%
\subsection{Hamiltonian formulation}
\label{subsec:hamiltonian_formulation}
%%%%%%%%%%%%%%%%%%%%%%%%%%%%%%%%%%%%

%%%%%%%%%%%%%%%%%%%%%%%%%%%%%%%%%
\subsubsection{Canonical momenta}
%%%%%%%%%%%%%%%%%%%%%%%%%%%%%%%%%

The canonical momentum conjugate to the scalar field $\phi$ is
\begin{equation}
\pi^\phi \define \pd{\Lagr}{\dot{\phi}}
=
\dot{\phi} \pa{ \pd{f_0}{\K} + \frac12\pd{f_1}{\K} \X + \frac12\pd{f_2}{\K} \Y - \pd{f_3}{\K} \Z}
+ f_3(\phi,\K) \partial_\mu \phi \tilde{F}\indices{^\mu^i}\tilde{F}\indices{^0_i} ,
\end{equation}
where an overdot stands for a time derivative $\dot{\phi}\define\partial_t\phi$; as usual, greek indices run from $0$ to $3$, while latin indices run from $1$ to $3$. The canonical momentum $\pi^\phi$ can be expressed in terms of the electric and magnetic parts $\vect{E}$, $\vect{B}$ of the field strength tensor, defined by
\begin{equation}
E^i \define F^{0i},
\qquad
B^i \define \tilde{F}^{0i} = \frac{1}{2} \eps^{ijk} F_{jk},
\end{equation}
(we use bold fonts for spatial vectors) as
\begin{multline}
\pi^\phi =
\dot{\phi} \paac{ \pd{f_0}{\K} 
						+ \pd{f_1}{\K} (\magn^2 - \elec^2) 
						- 2 \pd{f_2}{\K} \elec\cdot\magn 
						+ \pd{f_3}{\K} \pac{ (\magn\cdot\grad\phi)^2 - (\dot{\phi}\magn-\elec\times\grad\phi)^2 }
						} \\
+ f_3(\phi,\K) \pac{ \dot{\phi}\magn^2 + \det\pa{ \grad\phi,\elec,\magn } },
\end{multline}
where $\grad\define(\partial_i)$ is the spatial gradient, an in-line dot and a cross respectively denote the Euclidean scalar product $\vect{U}\cdot\vect{V}\define \delta_{ij}U^iV^j$ and vector product $(\vect{U}\times\vect{V})^k \define [ijk] U^iV^j$, and $\det(\vect{U},\vect{V},\vect{W}) \define (\vect{U}\times\vect{V})\cdot\vect{W} = [ijk] U^iV^jW^k$ is the 3-dimensional determinant.

The canonical momenta conjugate to the vector field $A^\mu$ are
\begin{align}
\pi^0 &\define \pd{\Lagr}{\dot{A}_0} = 0 \\
\pi^i &\define \pd{\Lagr}{\dot{A}_i}
= f_1(\phi,\K) F^{i0} 
+ f_2(\phi,\K) \tilde{F}^{i0} 
- f_3(\phi,\K) \eps^{ijk} \partial_k\phi \partial_\mu\phi \tilde{F}\indices{^\mu_j}.
\end{align}
When expressed in terms of the electric and magnetic fields, the
latter reads
\begin{equation}\label{eq:canonical_momentum_A}
\vect{\pi} 
=
 - f_1(\phi,\K) \elec
 -  f_2(\phi,\K) \magn
 - f_3(\phi,\K) \pac{ \dot{\phi}\magn\times\grad\phi - (\elec\times\grad\phi)\times\grad\phi }.
\end{equation}

%%%%%%%%%%%%%%%%%%%%%%%%%%%%%%%%%%%%%%%%%%%%%%%%%%%%%%%%%%%%%%
\subsubsection{Constraint on the nondynamical field component}
%%%%%%%%%%%%%%%%%%%%%%%%%%%%%%%%%%%%%%%%%%%%%%%%%%%%%%%%%%%%%%

Since the total Lagrangian density~$\Lagr$ does not involve any $\dot{A}_0$ term, $A_0$ is a nondynamical degree of freedom. The associated (Euler-Lagrange) equation of motion,
\begin{equation}
\pd{}{x^\mu} \pac{ \pd{\mathcal{L}}{(\partial_\mu A_0)} }
- \pd{\mathcal{L}}{A_0}
=0,
\label{eq:Euler_Lagrange_A0}
\end{equation}
is therefore a \emph{constraint}. Equation~\eqref{eq:Euler_Lagrange_A0} can be rewritten using $\partial\mathcal{L}/\partial{\dot{A}_0}=0$ and that $\partial_i A_0$ only appears within terms of the form $F_{i0}$, thus it always comes with $-\dot{A}_i$. As a consequence
\begin{equation}
\pd{\mathcal{L}}{(\partial_i A_0)}
=\pd{\mathcal{L}}{F_{i0}}
=-\pd{\mathcal{L}}{\dot{A}_i}
=-\pi^i,
\end{equation}
and the constraint reads
\begin{equation}\label{eq:constraint_A0}
\grad\cdot\vect{\pi} = 0.
\end{equation}
Had we considered terms breaking the gauge invariance in the action, then this constraint would have been altered on its right hand side. 

%%%%%%%%%%%%%%%%%%%%%%%%%%%%%%%%%%%
\subsubsection{Hamiltonian density}
%%%%%%%%%%%%%%%%%%%%%%%%%%%%%%%%%%%

Since the dark sector is decoupled from the other fields, its contribution to the Hamiltonian density is obtained by
\begin{equation}
\mathcal{H}_{\rm DS} \define \pi^\phi\dot{\phi} + \pi^i\dot{A}_i - \mathcal{L}_{\rm DS}.
\end{equation}
The canonical term $\pi^i\dot{A}_i$ can be rewritten in the following way:
\begin{equation}
\pi^i\dot{A}_i = \pi^i\pa{ F_{0i} + \partial_i A_0 }
				= \pi^i F_{0i} - A_0 \partial_i\pi^i 
					+ \partial_i\pa{ A_0 \pi^i },
\end{equation}
and the spatial divergence $\partial_i\pa{ A_0 \pi^i }$ can be dropped, since it would disappear in a boundary term while integrating $\mathcal{H}_{\rm DS}$ to build the Hamiltonian. Using the constraint \eqref{eq:constraint_A0} then yields
\begin{equation}
\pi^i\dot{A}_i 
= \pi^i F_{0i}
= - \vect{\pi} \cdot \elec.
\end{equation}
Finally, we inject the expression of the canonical momenta and of the Lagrangian density, and reorganize the various terms to obtain
\begin{equation}\label{eq:Hamiltonian_density}
\mathcal{H}_{\rm DS}
=
\sum_{a=0}^3 \mathcal{H}_a
\end{equation}
with
\begin{align}
\mathcal{H}_0 &= \frac12 f_0(\phi,\K) + \pd{f_0}{\K} \dot{\phi}^2, \\
\mathcal{H}_1 &= \frac12 f_1(\phi,\K) (\elec^2+\magn^2) + \pd{f_1}{\K} \dot{\phi}^2 (\magn^2-\elec^2), \\
\mathcal{H}_2 &= - 2 \pd{f_2}{\K} \dot{\phi}^2 \elec\cdot\magn, \\
\mathcal{H}_3 &= \frac12 f_3(\phi,\K) \pac{ (\magn\cdot\grad\phi)^2 + (\dot{\phi}\magn-\elec\times\grad\phi)^2 }
+ \pd{f_3}{\K} \dot{\phi}^2 \pac{ (\magn\cdot\grad\phi)^2 - (\dot{\phi}\magn-\elec\times\grad\phi)^2  }.
\end{align}
%
%\begin{multline}\label{eq:Hamiltonian_density}
%\mathcal{H}
%=
%\frac12 f_0(\phi,\K) + \pd{f_0}{\K} \dot{\phi}^2
%+ \frac12 f_1(\phi,\K) (\elec^2+\magn^2) + \pd{f_1}{\K} \dot{\phi}^2 (\magn^2-\elec^2)
%- 2 \pd{f_2}{\K} \dot{\phi}^2 \elec\cdot\magn \\
%+ \frac12 f_3(\phi,\K) \pac{ (\magn\cdot\grad\phi)^2 + (\dot{\phi}\magn+\elec\times\grad\phi)^2 }
%+ \pd{f_3}{\K} \pac{ (\magn\cdot\grad\phi)^2 - (\dot{\phi}\magn-\elec\times\grad\phi)^2  }.
%\end{multline}

In Eq.~\eqref{eq:Hamiltonian_density}, the Hamiltonian density is not expressed in terms of its natural variables, which are $\pi^\phi$, $\grad \phi$, $\vect{\pi}$, and $\grad A^\mu$. Here, we actually \emph{choose} to describe a physical state of the theory using the time derivatives of the fields instead of the canonical momenta. This is perfectly licit, since there exists a one-to-one and onto relation between both descriptions, and this choice will turn out to make the discussions of the following sections easier.

%Note also that Eq.~\eqref{eq:Hamiltonian_density} shall not be considered a perfectly explicit expression of the Hamiltonian density, because it still contains variables which are not independent from each others. Namely, $A_0$ is linked to $\vect{\pi}$ by the constraint \eqref{eq:constraint_A0}, itself linked to $\elec$, $\magn$ and $\phi$ by Eq.~\eqref{eq:canonical_momentum_A}.

%%%%%%%%%%%%%%%%%%%%%%%%%%%%%%%%%%%%%%%%%%%%%%%%
\subsection{Hamiltonian stability of the theory}
\label{subsec:hamiltonian_stability}
%%%%%%%%%%%%%%%%%%%%%%%%%%%%%%%%%%%%%%%%%%%%%%%%

In this subsection, we study the necessary conditions on the functions $f_{0,1,2,3}$ for the Hamiltonian density~\eqref{eq:Hamiltonian_density} to be bounded by below. Our method relies on proofs by contradiction: given some properties of $f_{0,1,2,3}$, we look for configurations of the fields $\phi, A^\mu$, such that $\mathcal{H}_{\rm DS}$ can be made arbitrarily negative. If at least one such state can be exhibited, then the theory is unstable, thus forbidden.
% if we insist that its quantum counterpart should be stable.

%%%%%%%%%%%%%%%%%%%%%%%%%%%%%%%%%%%%%%%%%%%%
\subsubsection{Conditions on $f_0$}
%%%%%%%%%%%%%%%%%%%%%%%%%%%%%%%%%%%%%%%%%%%%

For this paragraph only, and without loss of generality, we consider states for which $\elec=\magn=\vect{0}$, so that the contributions $\mathcal{H}_{1,2,3}$ of the Hamiltonian density do not enter into the discussion. There are two necessary conditions on $f_0$ for $\mathcal{H}_{\rm DS}$ to be bounded by below, namely:
\begin{description}
\item[$\boldsymbol{\partial f_0/\partial \K \geq 0}$.] \emph{If there existed a state $(\phi, \K)$ of the scalar field so that $\partial f_0/\partial \K < 0$}, then we could take $\dot{\phi}, |\grad\phi| \rightarrow\infty$ while keeping $\K$ constant, which would make the Hamiltonian density diverge towards $-\infty$. Such a situation is therefore excluded.
\item[$\boldsymbol{f_0(\phi,\K\geq 0)}$ must be bounded by below.] \emph{If there existed a positive value of $\K$ so that $\phi \mapsto f_0(\phi,\K)$ was not bounded by below}, then we could set the derivatives of $\phi$ so that $\dot{\phi}=0$, hence $\mathcal{H}_{\rm DS}=f_0/2$, which would not be bounded by below.
\end{description}

Note that the above reasoning does not apply for negative values of $\K$, since the term $\dot{\phi}^2 \partial f_0/\partial\K$ can possibly compensate the divergence of $f_0$. As an example, $f_0(\phi,\K)=\phi^2 \K$ is clearly not bounded by below for $\K<0$, but its contribution in the Hamiltonian density is
\begin{equation}
\mathcal{H}_0 = \frac{f_0}{2} + \pd{f_0}{\K} \dot{\phi}^2 = \frac{\phi^2}{2}\pac{ \dot{\phi}^2 + (\grad\phi)^2 } \geq 0,
\end{equation}
hence completely admissible.

%%%%%%%%%%%%%%%%%%%%%%%%%%%%%%%%%%%%%%%%%%%%
\subsubsection{Conditions on $f_3$}
%%%%%%%%%%%%%%%%%%%%%%%%%%%%%%%%%%%%%%%%%%%%

There are two conditions on $f_3$ for the Hamiltonian density to be bounded by below:
\begin{description}
\item[$\boldsymbol{\partial f_3/\partial\K=0}$.] \emph{If there existed a configuration~$(\phi,\K)$ so that this derivative was not zero}, then we could always tune $\elec$, $\magn$ so that $\partial_K f_3 [(\magn\cdot\grad\phi)^2 - (\dot{\phi}\magn-\elec\times\grad\phi)^2]<0$, and take $\dot{\phi}, |\grad\phi|\rightarrow\infty$ (while keeping $\K$ constant) which would imply $\mathcal{H}_{\rm DS}\rightarrow -\infty$. This divergence would have no chance to be compensated by the terms $\mathcal{H}_{0,1,2}$, since they are quadratic in the derivatives of $\phi$, whereas $\mathcal{H}_3$ is quartic.
\item[$\boldsymbol{f_3 \geq 0}$.] Consider a state for which $\elec\perp\magn$ and $\elec^2=\magn^2$, so that both $\mathcal{H}_2$ and the term associated with $\partial f_1/\partial\K$ vanish. Also set, for instance, $\grad\phi$ parallel to $\magn$, so that all the terms of $\mathcal{H}_{\rm DS}$ involving the electric and magnetic fields gather into
\begin{equation}
\pac{f_1 + \frac{f_3}{2} (\dot{\phi}^2+|\grad\phi|^2)} \magn^2 .
\label{eq:f3_positive}
\end{equation}
Thus, \emph{if there existed a configuration $(\phi,\K)$ so that $f_3(\phi,\K)<0$}, then the prefactor of $\magn^2$ in Eq.~\eqref{eq:f3_positive} could be made strictly negative by taking $\dot{\phi},|\grad\phi|$ large enough (while keeping $\K$ constant). Finally, $\magn^2\rightarrow\infty$ would imply $\mathcal{H}_{\rm DS}\rightarrow -\infty$.
\end{description}
Therefore, we consider $f_3(\phi,\K)=f_3(\phi) \geq 0$ from now on.

%%%%%%%%%%%%%%%%%%%%%%%%%%%%%%%%%%%%%%%%%%%%
\subsubsection{Conditions on $f_1$}
%%%%%%%%%%%%%%%%%%%%%%%%%%%%%%%%%%%%%%%%%%%%

The conditions on $f_1$ turn out to be the same as those on $f_3$, although their proofs are slightly subtler due to the difficulty of controlling the possible compensations between terms.
\begin{description}
\item[$\boldsymbol{\partial f_1/\partial\K=0}$.] \emph{If there existed a configuration $(\phi,\K)$ so that $\partial f_1/\partial\K>0$}, then a state with, for example, $\magn=\vect{0}$, $\elec$ parallel to $\grad\phi$, and $\dot{\phi}, |\grad\phi|\rightarrow\infty$ (while keeping $\K$ constant) would make $\mathcal{H}_{\rm DS}\rightarrow-\infty$.

\emph{If there existed a configuration $(\phi,\K)$ so that $\partial f_1/\partial\K<0$}, then we could choose a state where $\elec$, $\magn$, $\grad\phi$ are all orthogonal to each other, and
\begin{equation}
\frac{E}{B} = \frac{f_3 \sqrt{1+K/\dot{\phi}^2}}{f_3(1+K/\dot{\phi}^2)-2\partial_K f_1},
\end{equation}
so that
\begin{equation}
\mathcal{H}_{\rm DS}
=
\mathcal{H}_0 
+
\frac{f_1}{2} (\elec^2+\magn^2)
+
\underbrace{
\frac{f_3 \K/(2\dot{\phi}^2) - (\partial_\K f_1)^2}
		{f_3(1+\K/\dot{\phi}^2)/2 - \partial_\K f_1}
}_{\text{$<0$ for $\dot{\phi}$ large enough}}
\dot{\phi}^2 \magn^2 ;
\end{equation}
in this situation, $\dot{\phi}^2, \magn^2\rightarrow\infty$ (keeping $\K$ constant) would imply $\mathcal{H}_{\rm DS}\rightarrow-\infty$. 
\item[$\boldsymbol{f_1 \geq 0}$.] Consider a state for which $\dot{\phi}\magn=\elec\times\grad\phi$, so that $\mathcal{H}_2=\mathcal{H}_3=0$. \emph{If there existed a configuration $(\phi,\K)$ so that $f_1(\phi,\K)<0$}, then taking $\elec^2\rightarrow\infty$ or $\magn^2\rightarrow\infty$ would make $\mathcal{H}_{\rm DS}\rightarrow-\infty$.
\end{description}
Therefore, we consider $f_1(\phi,\K)=f_1(\phi) \geq 0$ from now on.

%%%%%%%%%%%%%%%%%%%%%%%%%%%%%%%%%%%%%%%%%%%%
\subsubsection{Condition on $f_2$}
%%%%%%%%%%%%%%%%%%%%%%%%%%%%%%%%%%%%%%%%%%%%

Just as $f_{1,3}$, $f_2$ cannot depend on $\K$ for the theory to be stable. Indeed, \emph{if there existed a configuration $(\phi,\K)$ so that $\partial f_2/\partial \K \not= 0$}, then we could consider a state for which $\elec$, $\magn$, $\grad\phi$ are aligned, with $\mathrm{sgn}(\elec\cdot\magn)=\mathrm{sgn}(\partial_\K f_2)$, and set for instance
\begin{equation}
\frac{E}{B} = \frac{1+f_3}{2|\partial_\K f_2|}.
\end{equation}
In this situation, the Hamiltonian density would become
\begin{equation}
\mathcal{H}_{\rm DS} = \mathcal{H}_0 + \mathcal{H}_1 - \dot{\phi}^2\magn^2 + \frac{f_3 \K \magn^2}{2},
\end{equation}
so that $\dot{\phi}^2, \magn^2 \rightarrow\infty$, while keeping $K$ constant, would imply $\mathcal{H}_{\rm DS}\rightarrow-\infty$. Hence, we can consider $f_2(\phi,\K)=f_2(\phi)$ from now on. Note that, contrary to $f_{1,3}$, there is no restriction on the sign of $f_2$, since the function itself does not appear in the Hamiltonian.

%%%%%%%%%%%%%%%%%%%%%%%%%%
\subsection{Summary and discussion}
\label{subsec:summary_stability}
%%%%%%%%%%%%%%%%%%%%%%%%%%

We have proved that, among the various couplings between the scalar field and the vector fields, many bring uncompensated instabilities in the theory, by making the Hamiltonian unbounded by below. In the framework chosen in this article, the most general Lagrangian density for the dark sector leading to a \emph{stable} theory is
\begin{equation}
\Lagr_{\rm DS}
=
-\frac12 f_0(\phi,\K)
- \frac14 f_1(\phi) \X
- \frac14 f_2(\phi) \Y
+ \frac12 f_3(\phi) \Z,
\label{eq:action_stable}
\end{equation}
where $f_1$, $f_3$ are positive functions, $\phi\mapsto f_0(\phi,\K \geq 0)$ is bounded by below, and $\partial f_0/\partial\K\geq 0$.
%As we will see in the next section, the causality condition actually implies $f_3\leq 0$, so that finally $f_3=0$ is required for the theory to be both stable and causal.

So far, our analysis has been performed on a Minkowski spacetime. Nevertheless, \emph{our conclusions remain valid in the presence of gravity, thanks to the equivalence principle}. Indeed, the divergences underlined in the previous paragraphs are local properties, namely, they regard the Hamiltonian \emph{density} rather than the Hamiltonian itself. Suppose one wishes to perform the same study in an arbitrary spacetime. Then, in the vicinity of any event~E, one is free to work in a free-falling frame, where spacetime is locally Minkowskian, and thus where the above calculations are valid (modulo negligible gravitational tidal effects). In other words, in the vicinity of E, one could construct a configuration of the fields so that the Hamiltonian density is arbitrarily negative. Note that this reasoning would not be true if the fields were non-minimally coupled to gravity, or more generally in any scenario where the equivalence principle is not respected.

%%%%%%%%%%%%%%%%%%%%%%%%%%%%%%%%%%%%%%%%%%%%%%%%%%
%%%%%%%%%%%%%%%%%%%%%%%%%%%%%%%%%%%%%%%%%%%%%%%%%%
\section{Causality of the models}
\label{sec:causality}
%%%%%%%%%%%%%%%%%%%%%%%%%%%%%%%%%%%%%%%%%%%%%%%%%%
%%%%%%%%%%%%%%%%%%%%%%%%%%%%%%%%%%%%%%%%%%%%%%%%%%

A field theory is considered causal if it admits an unambiguous notion of time  evolution; any initial condition of the fields--- i.e., their state on a spacelike hypersurface---must generate a unique final state through the equations of motion. In other words, time evolution must be a well-posed Cauchy problem. This is equivalent to the mathematical statement that the equations of motion must be hyperbolic, that is, whose second-order part involve a differential operator $G^{\mu\nu} \partial_\mu\partial_\nu$ with signature $(-,+,+,+)$, where the $(-)$-direction is timelike with respect to spacetime's metric.

%%%%%%%%%%%%%%%%%%%%%%%%%%%%%%%%%%%%%%%%%%%%%%%%%%
\subsection{Equations of motion}
%%%%%%%%%%%%%%%%%%%%%%%%%%%%%%%%%%%%%%%%%%%%%%%%%%

Let us first determine the equations of motion induced by the action~\eqref{eq:action_stable}. On the one hand, stationarity with respect to variations of the scalar field implies
\begin{equation}
0 = \frac{\delta S}{\delta \phi} 
	= \partial_\mu \pa{ \pd{f_0}{\K} \partial^\mu\phi } 
		- f_3 \partial_\mu \pa{ \tilde{F}^{\mu\rho} \tilde{F}\indices{^\nu_\rho} \partial_\nu\phi}
		- \frac12 \pd{f_0}{\phi}
		- \frac14 f'_1 \X - \frac14 f'_2 \Y
		+ \frac12 f_3' \Z,
\label{eq:EOM_scalar}
\end{equation}
where primes in $f'_{1,2,3}$ denote derivatives with respect to $\phi$. On the other hand, stationarity of the action with respect to variations of the vector field implies
\begin{equation}
0 = \frac{\delta S}{\delta A_\sigma} 
= \partial_\mu \pac{ f_1 F^{\mu\sigma} + f_2 \tilde{F}^{\mu\sigma} 
					- f_3 \eps^{\alpha\rho\mu\sigma} \partial_\alpha\phi\partial_\beta\phi\tilde{F}\indices{^\beta_\rho} }.
\label{eq:EOM_vector}
\end{equation}

Equations~\eqref{eq:EOM_scalar} and \eqref{eq:EOM_vector} form a coupled system of second-order differential equations for $\phi$ and $A^\mu$, which can be formally written as
\begin{equation}
\underbrace{
\begin{bmatrix}
\mathcal{D}_\phi^\phi & \mathcal{D}^\phi_{\sigma'} \\
\mathcal{D}^\sigma_\phi & \mathcal{D}^\sigma_{\sigma'}
\end{bmatrix}
}_{\boldsymbol{\mathcal{D}}}
\begin{bmatrix}
\phi \\ A^{\sigma'}
\end{bmatrix}
=
\begin{bmatrix}
H^\phi (\phi,\partial\phi,\partial A) \\ H^\sigma (\phi,\partial\phi,\partial A)
\end{bmatrix},
\label{eq:EOM_formal}
\end{equation}
where the first line corresponds to Eq.~\eqref{eq:EOM_scalar} and the second line to Eq.~\eqref{eq:EOM_vector}. The matrix of operators denoted $\boldsymbol{\mathcal{D}}$ contains the second-order part of the equations of motion, while $[H^\phi,H^\sigma]$ contains the remaining part. Explicitly, we have
\begin{align}
\mathcal{D}_\phi^\phi &= \pd{f_0}{\K} \, \Box + \pac{ 2 \pd[2]{f_0}{\K} \partial^\mu\phi\partial^\nu\phi
																					- f_3(\phi) \tilde{F}\indices{^\mu^\rho}\tilde{F}\indices{^\nu_\rho}
																					} \partial_\mu \partial_\nu,
\\
\mathcal{D}^\sigma_\phi &= \eta^{\sigma\sigma'} \mathcal{D}^\phi_{\sigma'}
											=  - f_3(\phi) \eps^{\alpha\rho\mu\sigma} \partial_\alpha\phi 
													\tilde{F}\indices{^\nu_\rho} \partial_\mu\partial_\nu,
\label{eq:EOM_extra_diagonal}\\
\mathcal{D}^\sigma_{\sigma'} &= f_1(\phi) \pa{ \delta^{\sigma}_{\sigma'} \Box - \partial^\sigma\partial_{\sigma'} }
- f_3(\phi) \eps^{\alpha\rho\mu\sigma} \eps\indices{^\beta_\rho^\nu_{\sigma'}} 
			\partial_\alpha\phi\partial_\beta\phi \partial_\mu\partial_\nu. 
\end{align}
where $\Box\define\partial^\mu\partial_\mu$ denotes the d'Alembertian.

%%%%%%%%%%%%%%%%%%%%%%%%%%%%%%%%%%%%%%%%%%%%%%%%%%
\subsection{Diagonalizing the system of equations of motion}
%%%%%%%%%%%%%%%%%%%%%%%%%%%%%%%%%%%%%%%%%%%%%%%%%%

As it appears clearly in the expression~\eqref{eq:EOM_extra_diagonal}
of $\boldsymbol{\mathcal{D}}$, the presence of $f_3$ couples the
equations of motion of the scalar and vector fields even in their
second-order part. As a consequence, we cannot investigate their
hyperbolicity independently from each other; instead, we must consider
the whole system~\eqref{eq:EOM_formal},
diagonalize\footnote{Concretely, this diagonalization procedure is
  equivalent to finding new fields, combinations of $\phi$ and
  $A^\mu$, whose second-order part of the equations of motion are
  decoupled.} it, and study the hyperbolicity of each
``eigenequation''. In practice, we proceed by diagonalizing the
\emph{principal symbol}~$\boldsymbol{\sigma}_{\mathcal{D}}(p_\mu)$ of
the system, defined as the matrix-valued polynomial obtained from the
principal differential operator~$\boldsymbol{\mathcal{D}}$ by
replacing $\partial_\mu$ with an abstract variable $p_\mu$.

In the expression of $\boldsymbol{\sigma}_{\mathcal{D}}$, there naturally appear three vectors, namely $p^\mu$, $\partial^\mu\phi$, and $\pFtilde^\mu\define p_\alpha \tilde{F}^{\alpha\mu}$, from which we can construct an orthonormal tetrad~$(e_a)_{a=1\ldots 4}$; assuming that $p^\mu$ is not null-like, we define indeed
\begin{align}
\label{eq:tetrad_1}
e_1^\mu &\define p^\mu / \sqrt{|p^2|},\\
\label{eq:tetrad_2}
e_2^\mu &\define \partial_\perp^\mu\phi
				/\sqrt{|(\partial_\perp\phi)^2|}
\qquad \text{with} \qquad			
\partial_\perp^\mu\phi \define \partial^\mu\phi 
				- \frac{(p^\nu\partial_\nu\phi)p^\mu}{p^2},
\\
\label{eq:tetrad_3}
e_3^\mu &\define \pFtilde_\perp^\mu / \sqrt{|\pFtilde_\perp^2|}
\qquad \text{with} \qquad
\pFtilde_\perp^\mu \define \pFtilde^\mu - \frac{(\pFtilde_\nu\partial_\perp^\nu\phi)\partial_\perp^\mu\phi}{(\partial_\perp\phi)^2},
\\
\label{eq:tetrad_4}
e_4^\mu &\define 
\eps^{\alpha\beta\gamma\mu} e_{1\alpha} e_{2\beta} e_{3\gamma}.
\end{align}
the orthogonality between $e_1^\mu$ and $e_3^\mu$ being ensured by the antisymmetry of $\tilde{F}^{\mu\nu}$. Let us use these notations to rewrite the various contractions involved in the symbol~$\boldsymbol{\sigma}_{\mathcal{D}}$,
\begin{align}
\tilde{F}\indices{^\mu^\rho}\tilde{F}\indices{^\nu_\rho}
p_\mu p_\nu
&=
\pFtilde^2,
\\
\eps\indices{^\alpha^\rho^\nu_{\sigma'}} \partial_\alpha\phi \tilde{F}\indices{^\mu_\rho} p_\mu p_\nu
&=
\sqrt{\abs{p^2(\partial_\perp\phi)^2\pFtilde_\perp^2}} \, e_{4\sigma'},
\\
p^\sigma p_{\sigma'} 
&=
\abs{p^2} e_1^\sigma e_{1\sigma'} ,
\\
\eps^{\alpha\rho\mu\sigma} \eps\indices{^\beta_\rho^\nu_{\sigma'}} 
			\partial_\alpha\phi\partial_\beta\phi p_\mu p_\nu
&=
\abs{p^2(\partial_\perp\phi)^2}
\pa{ e_4^2 \, e_3^\sigma e_{3\sigma'} 
	+ e_3^2 \, e_4^\sigma e_{4\sigma'} }.
\end{align}
Since the above expressions exhibit projections over the tetrad vectors (such as $e_{1\sigma'} e_1^\sigma$), we expect the symbol to be much simpler if it is written in the tetrad basis\footnote{The odd ordering of the vectors is chosen for the blocks of the matrix~\eqref{eq:principal_symbol_tetrad_basis} to appear more clearly.}~$(e_4,e_1,e_2,e_3)$ instead of the coordinate basis $( \partial_\mu)_{\mu=0\ldots 3}$; and indeed the result is
\begin{equation}
\boldsymbol{\sigma}_{\mathcal{D}}
=
\begin{bmatrix}
\sigma^\phi_\phi & \sigma^\phi_4 & 0 & 0 & 0  \\
\sigma^4_\phi & \sigma^4_4 & 0 & 0 & 0 \\
0 & 0 & 0 & 0 & 0 \\
0 & 0 & 0 & f_1 p^2 & 0 \\
0 & 0 & 0 & 0 & f_1 p^2 + f_3 (\partial_\perp\phi)^2 p^2
\end{bmatrix}.
\label{eq:principal_symbol_tetrad_basis}
\end{equation}
with
\begin{align}
\sigma^\phi_\phi &= \pd{f_0}{\K}\, p^2 + 2\pd[2]{f_0}{\K}\,(p^\mu\partial_\mu\phi)^2 - f_3(\phi) \pFtilde^2, \\
\sigma^\phi_4 &= (e_4)^2 \sigma^4_\phi = -(e_4)^2 f_3(\phi) \sqrt{\abs{p^2(\partial_\perp\phi)^2\pFtilde_\perp^2}}, \\
\sigma^4_4 &= f_1(\phi) p^2 + f_3(\phi) (\partial_\perp\phi)^2 p^2.
\end{align}
The five eigenvalues of $\boldsymbol{\sigma}_{\mathcal{D}}$ are therefore $\lambda_1=0$, $\lambda_2=f_1 p^2$, $\lambda_3
= f_1 p^2+ f_3(\partial_\perp\phi)^2 p^2$,
%
%\begin{equation}
%\lambda_3
%= f_1 p^2+ f_3(\partial_\perp\phi)^2 p^2
%= \pa{f_1g^{\mu\nu} 
%+ f_3 \K g^{\mu\nu}
%- f_3 \partial^\mu\phi\partial^\nu\phi } p_\mu p_\nu,
%\label{eq:third_eigenvalue}
%\end{equation}
%%
and the two solutions~$(\lambda_0,\lambda_4)$ of the second-degree equation $(\sigma^\phi_\phi-\lambda)(\sigma^4_4-\lambda)=\sigma^\phi_4\sigma^4_\phi$, that is
\begin{equation}
\pac{ \pd{f_0}{\K} \, p^2 + 2 \pd[2]{f_0}{\K} (p\cdot\partial\phi)^2 - f_3 \pFtilde^2-\lambda} \pac{ f_1p^2+f_3 (\partial_\perp\phi)^2 p^2 - \lambda } 
+ f_3^2 p^2(\partial_\perp\phi)^2\pFtilde_\perp^2 = 0,
\label{eq:trinom}
\end{equation}
where we used that $e_1^2e_2^2e_3^2e_4^2=-1$, since an orthonormal
tetrad has only one timelike vector. 

If $p^\mu$ is null, the construction of the tetrad is slightly different. One can set, e.g., $e_1^\mu=p^\mu$, and $e_2^\mu=\partial^\mu\phi/(p^\nu\partial_\nu\phi) - (\partial\phi)^2p^\mu/2(p^\nu\partial_\nu\phi)^2$,
so that both $e_1^\mu$ and $e_2^\mu$ are null vectors, and $e_1^\mu e_{2\mu}=1$. The other two ones, $e_3^\mu$ and $e_4^\mu$ are defined similarly as before, except that $\pFtilde_\perp^\mu$ must be $\pFtilde^\mu-(e_2^\nu\pFtilde_\nu)p^\mu$. One can check that the expression of the symbol in the basis~$(e_4,e_1,e_2,e_3)$ is then the same as in Eq.~\eqref{eq:principal_symbol_tetrad_basis} with $p^2=0$.

In principle, the second-order differential operators involved in the eigenequations of motion are obtained from the eigenvalues of the principal symbol~$\boldsymbol{\sigma}_{\mathcal{D}}$ by using the correspondance $p_\mu\leftrightarrow\partial_\mu$. This can be directly achieved for $\lambda_{1,2,3}$, yielding the eigenoperators
\begin{align}
\mathcal{D}_1 &= 0,\\
\mathcal{D}_2 &= f_1(\phi) \Box,\\
\mathcal{D}_3 &= \pac{ f_1(\phi) + f_3(\phi) K } \Box - f_3(\phi) \partial^\mu\phi \partial^\nu\phi \partial_\mu \partial_\nu.
\end{align}
The fact that one operator is zero is not suprising, because it translates that one of the degrees of freedom of the vector field is non-dynamical. As solutions of Eq.~\eqref{eq:trinom}, the last two eigenvalues of the principal symbol generally involve square roots, and it is unclear how one should interpret them in terms of differential operators. In the case $f_3=0$, however, $\boldsymbol{\sigma}_{\mathcal{D}}$ as written in Eq.~\eqref{eq:principal_symbol_tetrad_basis} is already diagonal, and the remaining differential operators read
\begin{align}
\mathcal{D}_0^{[f_3=0]}  &= \pd{f_0}{\K} \, \Box 
												+ 2\pd[2]{f_0}{\K} \, \partial^\mu\phi\partial^\nu\phi \partial_\mu\partial_\nu,
\label{eq:D0} \\
\mathcal{D}_4^{[f_3=0]} &= \mathcal{D}_2 = f_1(\phi) \Box.
\end{align}
%

%%%%%%%%%%%%%%%%%%%%%%%%%%%%
\subsection{Hyperbolicity of the eigenequations}
%%%%%%%%%%%%%%%%%%%%%%%%%%%%

It turns out that the third eigenoperator~$\mathcal{D}_3$ is actually sufficient to rule out the $f_3$-term. Indeed, consider for instance a state with a purely homogeneous scalar field\footnote{This does not restrict the generality of our discussion. Indeed, just as for the stability analysis, it is sufficient to find one counterexample (here a particular state for which the equations of motion are not hyperbolic) to exclude a theory, provided it is considered fundamental.}, then
\begin{equation}
\mathcal{D}_3 = f_1(\phi) \Box - f_3(\phi) \dot{\phi}^2 \Delta,
\end{equation}
where $\Delta\define\partial^i\partial_i$ is the Laplacian. We know from the stability analysis that $f_3\geq 0$, but if there exists a value of $\phi$ so that $f_3(\phi)>0$, then for $\dot{\phi}$ large enough, $\mathcal{D}_3$ becomes elliptical. Therefore, it is \emph{necessary} to have $f_3=0$ for the theory to be both stable and causal.

For $f_3=0$, the eigenoperators $\mathcal{D}_{1\ldots 4}$ are all proportional to $\Box$, which is hyperbolic with a timelike $(-)$-direction, thus the causality requirement does not impose further constraints on $f_1$. Concerning $f_0$, additionnally to the condition $\partial f_0/\partial\K\geq 0$ imposed by the Hamiltonian stability requirement, we must also have
\begin{equation}
\pd{f_0}{\K} + 2\K \pd[2]{f_0}{\K} \geq 0.
\label{eq:hyperbolicity_condition_k-essence}
\end{equation}
for the eigenoperator~$\mathcal{D}_0$ of Eq.~\eqref{eq:D0} to be hyperbolic. We propose, in Appendix~\ref{app:hyperbolicity_k-essence}, a simple proof of the above condition, which is well known\footnote{Note, by the way, that the discussion about hyperbolicity in one of the first reference article~\cite{1969PhRv..182.1400A} is partially wrong. Indeed, the authors claim that the hyperbolicity conditions are (a) $\partial f_0/\partial\K>0$, and (b) $\partial^2 f_0/\partial\K^2 \geq 0$, which is not really the case: (a) is rather imposed by the stability condition, and (b) does not exist at all. They also mention Ineq.~\eqref{eq:hyperbolicity_condition_k-essence}, but as a condition which ``assures the stability of the Cauchy problem---that is, small changes in the Cauchy data cannot produce large changes in the solution arbitrarily close to the initial surface''.} in the the context of k-essence~\cite{1969PhRv..182.1400A,ArmendarizPicon:2000ah,Rendall:2005fv,Bruneton:2006yp,Babichev:2007dw}.

%%%%%%%%%%%%%%%%%%%%%%%%%%%%%%%
%%%%%%%%%%%%%%%%%%%%%%%%%%%%%%%
\section{Conclusion and further remarks}\label{rd}
%%%%%%%%%%%%%%%%%%%%%%%%%%%%%%%
%%%%%%%%%%%%%%%%%%%%%%%%%%%%%%%

We have derived necessary conditions for the stability and causality of models built from one scalar field and one vector field coupled to each other. Under the restrictions stated in Sec.~\ref{sec:action}, we showed that the most general action describing a stable vector-scalar dark sector, whose dynamics is ruled by hyperbolic equations of motion, reads
\begin{equation}
S_{\rm DS}= \int \dd^4 x \sqrt{-g} \pac{ -\frac12 f_0(\phi,\K) 
													- \frac14 f_1(\phi) F^{\mu\nu} F_{\mu\nu}
													- \frac14 f_2(\phi) F^{\mu\nu}\tilde{F}_{\mu\nu}
												},
\end{equation}
with $\K\define\partial^\mu\phi\partial_\mu\phi$, and where the coupling functions obey:
\begin{itemize}
\item $\partial f_0/\partial \K \geq 0$, and  $f_0(\phi,\K\geq 0)$ bounded by below (stability);
\item $f_1(\phi) \geq 0$ (stability);
\item $\partial f_0/\partial \K + 2 K\partial^2f_0/\partial\K^2 \geq 0$ (hyperbolicity).
\end{itemize}   
There are no further restrictions over the coupling function $f_2$.  It is remarkable that the class of models satisfying the assumptions of Sec.~\ref{sec:action} are so constrained by the basic principles of stability and causality. However, it is worth noting that the theories excluded by our analysis are really ruled out only if one considers them as \emph{fundamental}. If, on the contrary, they represent the effective behavior of a more fundamental but healthy theory, then the only requirement is a reasonable domain of stability and causality. By essence, the present work cannot draw any definite conclusion within the world of such effective theories.

Gauge invariance was a central assumption in our analysis. We shall mention that an
important issue with this property in vector-field models was
pointed out in Ref.~\cite{Deffayet:2013tca}, where the authors
considered a possible generalization of electromagnetism in Minkowski
spacetime, inspired from scalar Galileon theories. Their conclusion
came in the form of a ``no-go theorem" for generalized vector field
Galileons, which states that it is impossible to construct more
general theories than standard electromagnetism, because all possible
extensions following the Galileon construction procedure lead to
topological or boundary terms, and are thus nondynamical. In order to
escape this theorem, one can however build models with multicomponent
gauge-invariant vector fields, or couple the vector field with another
field, e.g., a scalar field as done in this article. The coupling of
different types of fields with non-trivial dynamics was addressed
earlier in Refs.~\cite{Deffayet:2010zh, Deffayet:2013lga}, while
Ref.~\cite{Zhou:2011ix} proposed a complete study of scalar Galileons
with gauge symmetries. 

One may then wonder what models can be built once the condition of
gauge invariance is removed. References~\cite{Tasinato:2014eka,
  Heisenberg:2014rta} have recently addressed the problem of
gauge-invariance breaking for single-vector-field models, in the
spirit of Galileon theories. These analyses conclude that, for some
particular combinations of the non-gauge invariant terms, ghost-like
instabilities disappear and it is possible to obtain a well-behaved
Galileon-type generalization of the Proca theory with three physical
propagating degrees of freedom.  In general, dropping gauge invariance in a vector-scalar theory leads to a system with more physical degrees of freedom, the dynamics of which can be governed by a huge variety of terms in the action, corresponding to all the possible contractions formed out of $A_\mu, \partial_\mu \phi, \partial_\mu A_\nu$, such as  $A^\mu A_\mu$, $ A^\mu \partial_\mu \phi$, $ \partial^\mu A_\mu $,  $A_\mu A_\nu F^{\mu \alpha} F\indices{^\nu_\alpha}$, $A_\mu \partial_\nu \phi F^{\mu \alpha} F\indices{^\nu_\alpha}$, $\partial_\mu A_\nu  F^{\mu \alpha} F\indices{^\nu_\alpha}$, etc. Given that the structure of each of these terms is very different, there is a priori no general procedure to deal with all of them together, so that one should probably perform a dedicated analysis of the stability and causality for each model. 
%This is obviously beyond the scope of the present work. 
Nevertheless, we can point out that some particularly interesting terms are often studied, and admit a simple analysis; for instance, vector potentials of the form $V(A^2)$, or couplings of the form $A^\mu \partial_\mu \phi$, $ \partial^\mu A_\mu$. A very specific class of models of the form $f(F^2) + V(A^2)$ was recently studied in Ref.~\cite{Golovnev:2013gpa}, and where shown to have hyperbolic equations of motion for some special regions of phase space.

%We restrict to the generalization of the case which turned out to be stable in the case of ${\rm U}(1)$ invariant vector field. In this case, the field strength tensor for gauge fields reads:
%\be
%F_{a}^{\mu \nu} \equiv \partial^{\mu} A^{\nu} - \partial^{\nu} A_{a}^{\mu} - g  C_{a}^{bc}A_{b}^{\mu} A_{c}^{\nu} 
%\ee
%where $g$ is a coupling constant of the group, $  C_{a}^{bc}$ are the structure constants of the group and the index $(a)$ is a Lie algebra index. The generalization of the Lagrangian for the non-Abelian case becomes: 
%\ba\label{2ndna}
%S &=& \int \dd^4 x \sqrt{-g}
%\left[
%\frac{R}{2\kappa^{2}} 
%- f_0(\phi^{I}, K)  
%-  \frac{1}{4}f_{1}(\phi^{I})F_{a}^{\mu \nu}F_{a \mu \nu}
%- \frac{1}{4}f_{2}(\phi^{I})F_{a}^{\mu \nu}\tilde{F}_{a \mu \nu}
%\right].
%\ea
%Here,  $K =  G_{IJ}\partial_{\mu}\phi^{I} \partial^{\mu}\phi^{J} $ is the kinetic term and $G_{IJ}$ is the scalar field's space metric, $I, J$ indices label the scalar fields. The dual of the field strength becomes $\tilde{F}_{a \mu \nu} = (\star F)_{a \mu\nu} = \frac{1}{2} \varepsilon_{\mu \nu \alpha \beta}F^{a \alpha \beta}$.

%Here there is an important difference however with respect to the $U(1)$ commutative case. In the $U(1)$ case we see that the cubic terms formed out of contracting the field strength and its dual is zero, but this situation changes here because there appear the two non vanishing combinations formed out of $F$ and $\tilde{F}$: $FFF$ and $FF\tilde{F}$ so that, if we go further than the quadratic Lagrangian, we will find the terms:

Finally, we should mention that our analysis admits a straightforward generalization to multiple vector
fields which are gauge invariant under a non-Abelian gauge
group. However, in the non-Abelian case we have an important
difference with respect to the $\mathrm{U}(1)$ Abelian case. As shown in
Subsec.~\ref{subsec:general_class_Lagrangian}, in the $\mathrm{U}(1)$ case any term of the action involving the vector can always be reduced to even powers of the Faraday tensor or its
dual, possibly contracted with the derivatives of the scalar
field, which can be further reduced to products of $X$, $Y$ and
$Z$. All the odd products of the Faraday tensor and its dual are
identically zero.  In presence of a non-Abelian gauge group, this is
no longer the case, and there appear other combinations which add non-trivial dynamics to the system. Among the lowest order terms, there appear for instance cubic combinations of the form $FFF$  and $FF\tilde{F}$:
\begin{equation}
S_{\rm cubic} = -\int \dd^4 x \sqrt{-g} \left[ f(\phi^{I}) C_{abc} F^{a \mu \rho}F\indices{^{b}_{\rho \nu}}F\indices{^c^\nu _\mu}
+ g(\phi^{I}) C_{abc} F^{a \mu \rho}F^{b}_{\rho \nu}\tilde{F}\indices{^c^\rho_\nu}
\right],
\end{equation}
where $  C_{a}^{bc}$ are the structure constants of the group and
$a,b,c$ are Lie algebra indices. These terms are consistent with gauge
symmetries and are dynamical. 
 %\begin{tbm} Why? How do you know?\end{tbm}.  
 The term  $FFF$ appears generically in non-Abelian gauge theories and terms like $FF\tilde{F}$ appear for instance in QCD when discussing   CP-violations originated by gluonic operators of dimension six  \cite{Weinberg:1989dx}. Recently, the dynamics of such terms was also studied in the context of leptogenesis in  non-Abelian gauge fields populated models \cite{Maleknejad:2014wsa}.   Despite of the numerous ways in which the the fields can interact when gauge invariance is broken or when non-Abelian gauge groups are considered,  it is expected that the stability and the causality analysis would impose constraints over all those possible interactions and it would be interesting and valuable to extend the methods followed here to those cases.

%{\color{red}{Some words about non-minimal coupling to gravity.}} In Ref. \cite{Jimenez:2013qsa} it was studied the vector-tensor interactions in the context of Horndeski theories. They found that superluminality problems appear generically in these theories which points problems with the hyperbolicity of the equations of motion of the theory. 

%To conclude...{\color{red}{we briefly summarize/emphasize in a couple of sentences our main conclusions.}}

%%%%%%%%%%%%%%%%%%%%%%%
\section*{Acknowledgments}
We thank Gilles Esposito-Far\`ese for useful discussions. We also thank C\'esar Valenzuela-Toledo and Yeinzon Rodr\'{i}guez for collaboration at early stages of this project, and Mikjel Thorsrud for relevant comments on the first version of this manuscript.
This work was partly supported by COLCIENCIAS --- ECOS NORD grant number RC 0899-2012 and by French state funds managed by the ANR within the Investissements d'Avenir programme under reference ANR-11-IDEX-0004-02. J.P.B.A. was supported by VCTI (UAN) grant number 20131041; he thanks Institut d'Astrophysique de Paris, and joins P.F. and C.P. to thank Universidad del Valle, for their warm hospitality and stimulating academic atmosphere during several stages of this project. The internal preprint number of this paper is PI/UAN-2014-571T.

%%%%%%%%%%%%%%%%%%%%%%% 

%%%%%%%%%%%%%%%%%%%%%%%%%%%%-Appendix-%%%%%%%%%%%%%%%%%%%%%%%%%%%%
\appendix

%%%%%%%%%%%%%%%%%%%%%%%%%%%%%%%
%%%%%%%%%%%%%%%%%%%%%%%%%%%%%%%
\section{Hyperbolicity of the scalar sector}
\label{app:hyperbolicity_k-essence}
%%%%%%%%%%%%%%%%%%%%%%%%%%%%%%%
%%%%%%%%%%%%%%%%%%%%%%%%%%%%%%%

Consider the differential operator
\begin{equation}
\pa{ \pd{f_0}{\K} \eta^{\mu\nu} + 2 \pd[2]{f_0}{\K} \partial^\mu\phi\partial^\nu\phi }
\partial_\mu\partial_\nu 
\define
G^{\mu\nu} \partial_\mu\partial_\nu .
\end{equation}
Its hyperbolicity can be investigated by distiguishing two cases, depending on the sign of $\K$.
\begin{enumerate}
\item $K<0$. Define $n^\mu\define-\partial^\mu\phi/\sqrt{-K}$. Since $n^\mu$ is a unit timelike vector, we can always find a Lorentz tranformation $\Lambda$ such that $n^\alpha=\Lambda\indices{^\alpha_\mu} n^\mu = \delta^\alpha_0$. Thus, in this new frame ($G^{\alpha\beta}\define \Lambda\indices{^\alpha_\mu}\Lambda\indices{^\beta_\nu} G^{\mu\nu}$), the differential operator reads
\begin{equation}
G^{\alpha\beta}\partial_\alpha\partial_\beta
= -\pa{ \pd{f_0}{\K} + 2 K \pd[2]{f_0}{\K} } \partial^2_0 + \pd{f_0}{\K} \Delta,
\end{equation}
and is therefore hyperbolic if and only if $\partial f_0/\partial\K + 2 K \partial^2 f_0/\partial\K^2 \geq 0$ (additionally to the stability condition $\partial f_0/\partial\K\geq 0$).
\item $K>0$. Define $n^\mu\define \partial^\mu\phi/\sqrt{K}$, which is now a unit spacelike vector, thus there exists a Lorentz transformation $\Lambda$ such that, e.g., $n^\alpha=\Lambda\indices{^\alpha_\mu} n^\mu = \delta^\alpha_1$. In this new frame, the differential operator becomes
\begin{equation}
G^{\alpha\beta}\partial_\alpha\partial_\beta
= - \pd{f_0}{\K} \partial_0^2 + \pa{ \pd{f_0}{\K} + 2 K \pd[2]{f_0}{\K} } \partial^2_1 + \pd{f_0}{\K} \Delta_{\rm 2D},
\end{equation}
and is hyperbolic under the same condition as in the case $K<0$ above.
\item $\K=0$. This case is slightly less trivial than the previous two ones. Up to a spatial rotation, we can write $\partial^\mu\phi=\dot{\phi}(-\delta_t^\mu+\delta_1^\mu)$. The differential operator $G^{\mu\nu}\partial_\mu\partial_\nu$ can then be diagonalized using the two vectors
\begin{equation}
\partial_\pm \define 
\frac{f_0''\dot{\phi}^2 \partial_0 - \pac{ f_0' \pm \sqrt{ (f_0')^2 + (f''_0\dot{\phi}^2)^2 } }\partial_1}
		{ \sqrt{ (f_0''\dot{\phi}^2)^2 + \pac{ f_0' \pm \sqrt{ (f_0')^2 + (f''_0\dot{\phi}^2)^2 } }^2  } } ,
\end{equation}
where we denoted $f_0'\define\partial f_0/\partial\K$ for short. Using the basis $(\partial_-,\partial_+,\partial_2,\partial_3)$, the differential operator indeed becomes
\begin{equation}
G^{\alpha\beta}\partial_\alpha\partial_\beta = G^{-} \partial_-^2 + G^{+} \partial_+^2 + \Delta_{2\rm D},
\end{equation}
with
\begin{align}
G^{-} &\define -f_0''\dot{\phi}^2 - \sqrt{ (f_0')^2 + (f''_0\dot{\phi}^2)^2 } \leq 0,\\
G^{+} &\define -f_0''\dot{\phi}^2 + \sqrt{ (f_0')^2 + (f''_0\dot{\phi}^2)^2 } \geq 0.
\end{align}
Thus, the differential operator is always hyperbolic. The question is now whether the $(-)$-direction is timelike or spacelike; it is immediate to check that
\begin{equation}
g(\partial_\pm,\partial_\pm) = \frac{\pm 2 f_0' \sqrt{ (f_0')^2 + (f''_0\dot{\phi}^2)^2 }}
												{(f_0''\dot{\phi}^2)^2 + \pac{ f_0' \pm \sqrt{ (f_0')^2 + (f''_0\dot{\phi}^2)^2 } }^2},
\end{equation}
so that $\partial_-$ is timelike (and $\partial_+$ is spacelike) if and only if $f_0'=\partial f_0/\partial\K\geq 0$, which is consistent with the condition found in the previous two cases, for $\K=0$.
\end{enumerate}

%%%%%%%%%%%%%%%%%%%%%%%%%%%%%%%
%%%%%%%%%%%%%%%%%%%%%%%%%%%%%%%
\section{Beyond linearity in $\X$, $\Y$, $\Z$}
\label{app:beyond_linearity_X_Y_Z}
%%%%%%%%%%%%%%%%%%%%%%%%%%%%%%%
%%%%%%%%%%%%%%%%%%%%%%%%%%%%%%%

In this appendix, we provide some results that can be a starting point for the analysis of more general models than the ones described by action~\eqref{eq:action}. Namely, consider a dark-sector Lagrangian density $\mathcal{L}_{\rm DS}(\phi,\K,\X,\Y,\Z)$ which is not necessarily linear in $\X=F^2$, $\Y=F\tilde{F}$, and $\Z=(F\partial\phi)^2$. In the following, we drop the `DS' label  to alleviate notation.

%%%%%%%%%%%%%%%%%%%%%%%%%%%%%%%
\subsection{Hamiltonian density}
%%%%%%%%%%%%%%%%%%%%%%%%%%%%%%%

The canonical momenta conjugate to the scalar and vector fields are respectively
\begin{align}
\pi^\phi &= -2\Lagr_{,\K} \dot{\phi} + 2 \Lagr_{,\Z} \pac{ \dot{\phi} \magn^2 + \det(\grad\phi,\elec,\magn) },\\
\vect{\pi} &= 4 \Lagr_{,\X} \elec + 4 \Lagr_{,\Y} \magn 
					- 2 \Lagr_{,\Z}\pac{ \dot{\phi}\magn\times\grad\phi - (\elec\times\grad\phi)\times\grad\phi },
\end{align}
where a coma denotes a derivative. The constraint is unchanged compared to the case where $\Lagr$ is linear in $\X$, $\Y$, $\Z$, namely $\grad\cdot\vect{\pi}=0$. The Hamiltonian density then reads
\begin{equation}
\mathcal{H}_{\rm DS} = -2 \Lagr_{,\K} \dot{\phi}^2 - 4 \Lagr_{,\X} \elec^2 + \Y \Lagr_{,\Y} 
						+ 2 \Lagr_{,\Z} (\dot{\phi} \magn - \elec\times\grad\phi)^2 - \Lagr(\K,\X,\Y,\Z),
\end{equation}
and $\K$, $\X$, $\Y$, and $\Z$ are expressed in terms of the fields as
\begin{align}
\K &= (\grad\phi)^2 - \dot{\phi}^2, \\
\X &= 2 (\magn^2 -\elec^2), \\
\Y &= -4 \elec\cdot\magn,\\
\Z &= (\dot{\phi}\magn - \elec\times\grad\phi)^2 - (\magn\cdot\grad\phi)^2.
\end{align}
It would be tempting to conclude that $\Lagr_{,K}\, \Lagr_{,\X}\leq 0$, and $\Lagr_{,\Z}\geq 0$ are necessary conditions for $\mathcal{H}_{\rm DS}$ to be bounded by below, but unfortunately $\dot{\phi}^2$, $\elec^2$, $(\dot{\phi} \magn - \elec\times\grad\phi)^2$, $\K$, $\X$, $\Y$, $\Z$ are not independent variables, so that one cannot take, e.g., $\dot{\phi}^2\rightarrow\infty$ while keeping the others finite. The actual stability conditions could be much subtler, for instance they could involve combinations of the derivatives of $\Lagr$, and thus require a dedicated study.

%%%%%%%%%%%%%%%%%%%%%%%%%%%%%%%
\subsection{Equations of motion}
%%%%%%%%%%%%%%%%%%%%%%%%%%%%%%%

The equations of motion induced by the general Lagrangian~\eqref{eq:general_Lagrangian_DS} read
\begin{align}
0 &= \frac{\delta S}{\delta\phi} 
	= \Lagr_{,\phi} - \partial_\mu \pa{ 2\Lagr_{,\K}\partial^\mu\phi 
															+ 2\Lagr_{,\Z} \tilde{F}^{\mu\alpha}\tilde{F}\indices{^\nu_\alpha}\partial_\nu\phi
															}, \\
0 &= \frac{\delta S}{\delta A_\sigma} 
	= -\partial_\mu\pa{ 4 \Lagr_{,\X} F^{\mu\sigma} 
									+ 4 \Lagr_{,\Y} \tilde{F}^{\mu\sigma} 
									+ 2 \Lagr_{,\Z} \eps^{\beta\gamma\mu\sigma} \partial_\alpha\phi\partial_\beta\phi 
														\tilde{F}\indices{^\alpha_\gamma} 
									}.
\end{align}
As in Eq.~\eqref{eq:EOM_formal}, we can isolate the second order part of the above system, and write it as the matrix-valued differential operator
\begin{equation}
\boldsymbol{\mathcal{D}}
\define
\begin{bmatrix}
\mathcal{D}_\phi^\phi & \mathcal{D}^\phi_{\sigma'} \\
\mathcal{D}^\sigma_\phi & \mathcal{D}^\sigma_{\sigma'}
\end{bmatrix},
\end{equation}
where
\begin{multline}
\mathcal{D}_\phi^\phi = \Big( \Lagr_{,K} \eta^{\mu\nu} 
												+ \Lagr_{,Z} \tilde{F}^{\mu\rho} \tilde{F}\indices{^\nu_\rho}
												+ 2 \Lagr_{,KK} \phi^{,\mu} \phi^{,\nu}
												+ 4 \Lagr_{,KZ} \phi^{,\mu} \phi_{,\alpha} \tilde{F}^{\alpha\rho}\tilde{F}\indices{^\nu_\rho} \\
												+ 2 \Lagr_{,ZZ}  \phi_{,\alpha} \phi_{,\beta} \tilde{F}^{\mu\rho} \tilde{F}\indices{^\alpha_\rho}
																														\tilde{F}^{\nu\sigma} \tilde{F}\indices{^\beta_\sigma}
												\Big) \partial_\mu\partial_\nu,
\end{multline}
\begin{multline}
\mathcal{D}^{\sigma}_\phi = \eta^{\sigma\sigma'} \mathcal{D}_{\sigma'}^\phi 
											= \Big[ \Lagr_{,Z} \eps^{\alpha\rho\mu\sigma}\phi_{,\alpha} \tilde{F}\indices{^\nu_\rho}
												 		+ 4 \phi^{,\mu} \pa{ \Lagr_{,KX} F^{\nu\sigma} 
												 										+  \Lagr_{,KY} \tilde{F}^{\nu\sigma} }
												 		+ 2 \Lagr_{,KZ} \eps^{\alpha\rho\mu\sigma} 
												 									\phi^{,\nu} \phi_{,\alpha}\phi_{,\beta} \tilde{F}\indices{^\beta_\rho} \\
												 		+ 4 \phi_{,\alpha} \tilde{F}^{\mu\rho}\tilde{F}\indices{^\alpha_\rho}
												 				\pa{ \Lagr_{,ZX} F^{\nu\sigma}	+ \Lagr_{,ZY}\tilde{F}^{\nu\sigma} }
												 		+ 2 \Lagr_{,ZZ} \eps^{\alpha\rho\mu\sigma} \phi_{,\alpha}\phi_{,\beta}\phi_{,\gamma}
												 			\tilde{F}\indices{^\beta_\rho}\tilde{F}^{\nu\lambda}\tilde{F}\indices{^\gamma_\lambda}
												 \Big] \partial_\mu\partial_\nu,
\end{multline}
\begin{multline}
\mathcal{D}^{\sigma\sigma'} = \Big\{ 2 \Lagr_{,X} \pa{ \eta^{\mu\nu}\eta^{\sigma\sigma'} 
																							- \eta^{\mu\sigma}\eta^{\mu\sigma'} }
															+ \Lagr_{,Z} \eps^{\alpha\rho\mu\sigma}\eps\indices{^\beta_\rho^\nu^{\sigma'}}
																					\phi_{,\alpha} \phi_{,\beta}
															+ 8 \Lagr_{,XX} F^{\mu\sigma}F\indices{^\nu^{\sigma'}} \\
															+16 \Lagr_{,XY} F^{\mu(\sigma}\tilde{F}\indices{^\nu^{\sigma')}}
															+ 8 \Lagr_{,YY} \tilde{F}^{\mu\sigma} \tilde{F}\indices{^\nu^{\sigma'}}
															+ 8 \eps\indices{^\alpha^\rho^\mu^{(\sigma}} \phi_{,\alpha}\phi_{,\beta}
																	\pac{ \Lagr_{,XZ} F^{\nu\sigma')}F\indices{^\beta_\rho} 
																			+ \Lagr_{,YZ} F^{\nu\sigma')}\tilde{F}\indices{^\beta_\rho} } \\
															+ 2 \Lagr_{,ZZ} \eps^{\alpha\rho\mu\sigma}\eps\indices{^\beta^\lambda^\nu^{\sigma'}}
																						\phi_{,\alpha}\phi_{,\beta}\phi_{,\gamma}\phi_{,\delta}
																						\tilde{F}\indices{^\gamma_\rho} \tilde{F}\indices{^\delta_\lambda}
												 \Big\} \partial_\mu\partial_\nu,
\end{multline}
with the symmetrization convention $T^{(\mu\nu)}\define (T^{\mu\nu}+T^{\nu\mu})/2$. The above formulae can be used for investigating the hyperbolicity of the equations of motion.

\bibliographystyle{JHEP} 
\bibliography{Biblio_vector_fields}

\end{document}